# MEASUREMENT OF THE VORTEX SPECTRUM IN A VORTEX-BEAM ARRAY WITHOUT CUTS AND GLUING OF THE WAVEFRONT

A. Volyar, M. Bretsko, Ya. Akimova, Yu. Egorov

*V.I. Vernadsky Crimean Federal University, Vernadsky Prospekt, 4, Simferopol, 295007, Russia*

**Abstract**

We presented a new method for measuring the squares of the amplitudes and phases of partial vortex-beams in a complex beam array in real time. The method is based on measuring the high-order intensity moments and analyzing the solutions of a system of linear equations. Calibration measurements have shown that the measurement error at least for an array of 10-15 beams does not exceed 4%. When measuring beams with different fractional topological charge of optical vortices, we revealed a sharp decrease in the OAM for small deviations of the topological charge from the integer order which is associated with a sharp increase in the number of integer-order vortices. The application of this approach e.g. for the terabit-information processing in optical fibers provides significant advantages in comparison with the traditional diffraction methods and opens up additional possibilities of information compressor when modulating not only the OAM but also phases of partial beams

The intensive development of singular optics stimulates the search of new express methods for creating and measuring the composition of optical vortices (the vortex spectrum) and the orbital angular momentum (OAM) in complex arrays of simple vortex-beams. The term "vortex spectrum" as used in the text means the dependence of the square of the amplitude and phase of the standard vortex-beams on their azimuthal index (topological charge). In its physical nature, an optical vortex is a simple phase structure with a helix wavefront, whereas an array of optical vortices forms a peculiar phase skeleton of a complex paraxial beam [1]. Even a weak effect on the individual vortices of the array leads to a distortion of the phase skeleton and, consequently, to the loss of basic information embedded in the singular beam. There are many approaches to measuring the properties of vortices in an array based on aperturing some sections of the wavefront surface (Shack–Hartman method) or sorting individual beams in accordance with their azimuthal index (see, e.g. a detailed review in [1] and references therein). All of them are accompanied by cuts of the wavefront and partial loss of information. Nevertheless, important properties of the optical vortex array can be preserved and improved provided that diffraction optical elements (DOEs) are used for telecommunications systems, cryptography, entanglement devises, etc [2-5]. The first attempts to transmit compressed information flows in complex beams due to the coding of OAM via DOEs were associated with a low quality of the optical vortex separation and the overlapping of various partial beams in the same diffraction orders [6]. A significant contribution to the technology of DOEs was made by Kotlyar, Khonina, Soifer et al (see Ref. [7-10].and references therein). Their holographic filters made it possible to form complex beams and to separate a great number of vortex-beams comprising in an optical array. Further development of this approach was implemented in a variety of devices for terabit information transmission through free space or optical fibers [11,12]. However, many optical wave structures, such as vortex-beams with a fractional topological charges, optical quarks, etc [13-15] require continuous monitoring of the amplitude, phase, polarization, and OAM of the partial vortices in the wave array, which can not provide diffraction methods

Is it possible in principle to simultaneously measure the amplitude, phase, and OAM of partial vortices in a complex beam array without destroying its internal structure? The use of refractive elements to measure OAM [16] does not solve the problem, since it introduces significant phase and amplitude distortions into the beam. The vortex spectrometer proposed in [17] also does not provide a solution to the problem, since a change in the beam aperture and its subsequent coupling with a single-mode fiber breaks substantially the structure of the vortex array.

In connection with this, we focused attention on the properties of the Wigner function distribution and its adaptation by Hu for optics in the form of intensity moments [18]. The fact is that the second-order intensity moments make it possible to measure not only certain parameters of a paraxial beam [19,20], but also to control the inner state of optical vortices (intrinsic and extrinsic OAM etc) without changing the beam structure [21-23].

The aim of our letter is to demonstrate ant to test a method for controlling the vortex spectrum of a vortex-beam array via a measurement of high-order intensity moments

**1.** As the simplest model for analysis, we choose a scalar wave field in the form of a superposition of the $N$ monochromatic Laguerre-Gauissian beams $LG_n^m$ of the lowest order, where $n$ and $m$ are the radial and azimuthal indices, respectively, so that all the beams have the same waist radius $w_0$ at $z = 0$. We shall consider the wave field in the waist plane in the form

$$\Psi(r,\varphi,z=0) = \sum_{m=0}^{N-1} C_m LG_{n=0}^m = \sum_{m=0}^{N-1} \frac{C_m r^m e^{i(m\varphi+\beta_m)} G(r)}{N_m}, \quad (1)$$

where $G = \exp(-r^2)$, $\varphi-$ is an azimuthal angle, $r = \sqrt{x^2+y^2}/w_0$ $r$ is normalized radial coordinate, $N_m = \sqrt{2^{-m-1} m! \pi}$ stands for the normalization factor, $C_m$ is a beam amplitude, $\beta_m$ is the initial phase. Generally speaking, the representation of a complex beam field can be chosen for any basis of orthogonal functions. However, we chose the most appropriate version (1), acceptable for reading and understanding. Besides, we consider the nondegenerate case of the field representation $(m \geq 0 \text{ or } m \leq 0)$, since the intensity moments do not distinguish beams with axial symmetry but different signs of topological charges $\pm m$.

The intensity distribution of the beam can be represented in the form

$$\Im = \Psi^*\Psi = \sum_{m=0}^{N-1} \frac{C_m^2}{N_m^2} r^{2m} G^2 + 2 \sum_{\substack{m,n=0,\\m\neq n}}^{N-1} \frac{C_n C_m}{N_m N_m} r^{m+n} \cos[(m-n)\varphi] \cos\beta_{m,n} G^2 -$$
$$-2 \sum_{\substack{m,n=0,\\m\neq n}}^{N-1} \frac{C_n C_m}{N_m N_m} r^{m+m} \sin[(m-n)\varphi] \sin\beta_{m,n} G^2 \quad (2)$$

Note that the first term in the expression (2) depends only on the squares of the amplitudes $C_b^2$, the second term contains only $\cos\beta_{m,n}$, $\beta_{m,n} = \beta_m - \beta_n$ and the third - only $\sin\beta_{m,n}$. Our task is to measure each term of all sum. To this end, we introduce the standard expression for the intensity moments [18]

$$J_{p,q} = \int_{-\infty}^{\infty}\int_{-\infty}^{\infty} x^p y^q \Im(x,y) dx dy / J_{0,0}, \quad (3)$$

where $p,q = 0, 1, 2, 3,,,$, $J_{0,0} = \sum_{m=0}^{N-1} C_m^2$ is the beam intensity.

The intensity moments (3) can be experimentally measured [19]. Now the problem is to choose a combination of intensity moments $J_{p,q}$ in such a way as to exclude all terms of the last two sums and leave only the first one in eq. (3). A direct calculation shows that this requirement is satisfied by the relation

$$J_{2p} = \sum_{j=0}^{p} \binom{p}{j} J_{2j,2(p-j)}, \quad p = 1, 2, 3..., N-2. \quad (4)$$

Substituting eq. (2) into eq. (3) and taking into account relation (4), we obtain a system of $N-2$ linear equations for the squares of amplitudes $C_m^2$

$$\sum_{m=0}^{N-1}\frac{(N-1+m)!}{m!}C_m^2 = \sum_{j=0}^{p}\binom{p}{j}J_{2j,2(p-j)}, \quad p \leq N-2. \quad (5)$$

In order that the system of equations become complete, we put formally $p=1/2$ then we find

$$\sum_{m=0}^{N-1}\frac{m+1/2}{4\sqrt{2}\,m!}\Gamma(m+1/2)C_m^2 = \sqrt{J_{2,0}+J_{0,2}}. \quad (6)$$

Thus, the right-hand sides of the equations are measured experimentally, then the solution of the linear system (5)-)6) gives the desired values of squares of the amplitudes $C_m^2$.

Since the intensity of the beam array depends only on the phase difference between two pairs of modes, in calculating the initial phases $\beta_m$ we will assume that the phase of one of the beams is given (say $\beta_0 = 0$). Now it is necessary to choose the combination of intensity moments $J_{p,q}$ so that either the second or the third sum in eq.(2) is taken into account in the calculation. It is natural to calculate not all the phase differences $\beta_{n,m}$, but only $\beta_{0,m}$. However, we did not find such combinations $J_{p,q}$ that could filter out only terms with $\beta_{0,m}$.

Generally speaking, to calculate the phase difference $\beta_{m,n}$ it is required to find the $M$ equations for variables $X_{m,n} = C_m C_n \cos\beta_{m,n}$ and $Y_{m,n} = C_m C_n \sin\beta_{m,n}$ in the second and third sum in eq.(2) whose number is equal to the number of 2-combinations from $N$ elements.

It turns out that the number of equations can be significantly reduced if we use the moments $J_{2(p+1),1}$ for $X_m$ variables and $J_{1,2(p+1)}$ for $Y_m$ variables. The system of linear equations for the phase difference are written in the form

$$J_{2(p+1),1} = \frac{1}{2^{2(p+1)}}\sum_{m=0,\,k=0}^{N-1}\sum^{p}\binom{2p}{k}\left[\frac{M_{m,n_1}}{N_n N_{n_1}}X_{m,n_1}+\frac{M_{m,n_2}}{N_m N_{n_2}}X_{m,n_2}\right], \quad (7)$$

$$J_{1,2(p+1)} = \frac{1}{2^{2(p+1)}}\sum_{m=0,\,k=0}^{N-1}\sum^{p}(-1)^{p-k}\binom{2p}{k}\left[\frac{M_{m,n_1}}{N_m N_{n_1}}Y_{m,n_1}+\frac{M_{m,n_2}}{N_m N_{n_2}}Y_{m,n_2}\right], \quad (8)$$

where $n_1 = |m-2(p-k)\pm 3|$, $n_1 = |m-2(p-k)\pm 1|$, $M_{m,n_{1,2}} = \int_0^\infty r^{m+n_{1,2}+1}G^2 dr$. The number of

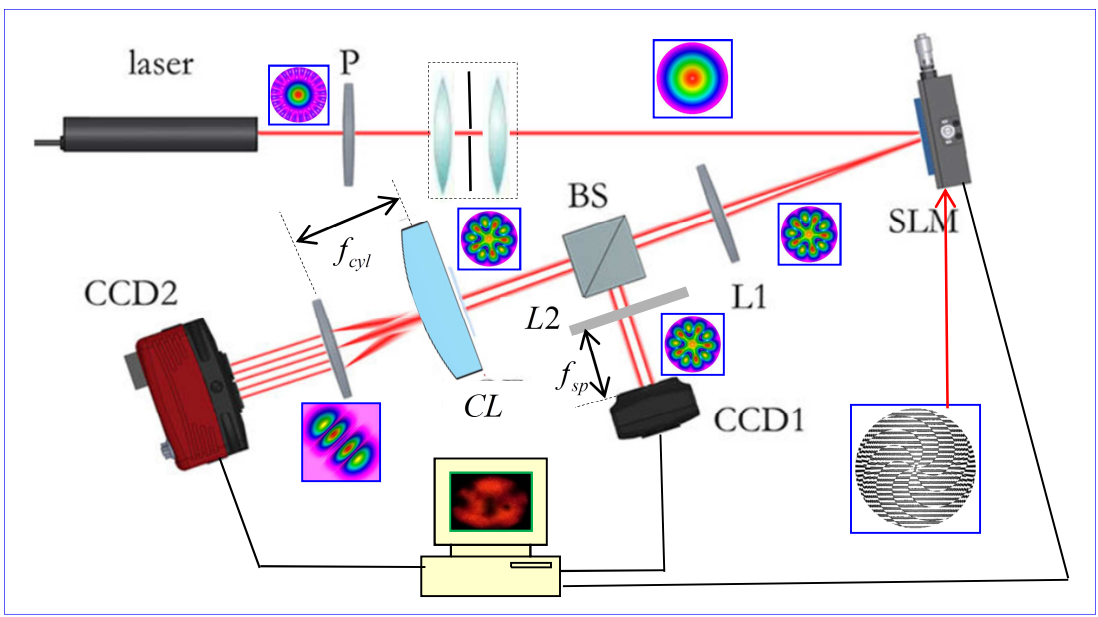

Fig.1 Exprimerimental setup for real-time measuring the vortex and the OAM spectrum, P – polarizer, FF – space light filter, SLM – space light modulator, L1, L2 – spherical lenses with a focal length $f_{sh}$, BS – beam splitter, CL – cylindrical lent with a focal length $f_{cyl}$, CCD1,2 – CCD camera.

equations in each system (7) and (8) is $K = 3(N-3)$, $N \geq 6$. It is noteworthy that the eqs (7) and (8) contain only terms with an odd difference of indices, including $\beta_{0,m}$, so that a finite solution enables us to obtain all phases of the partial beams in the form $\tan \beta_{m,n} = Y_{m,n} / X_{m,n}$.

**2.** Measurements of the square of the amplitude, phase, and the OAM of the array of $N$ beams were carried out on the experimental setup in Fig.1. The laser beam $\lambda = 0.628\,mcm$ passes through a spatial filter SF and is projected onto a spatial light modulator SLM that forms an array of $N$ singular beams with given amplitudes $C_m$ and phases $\beta_m$. The reflected beam is split into two arms by a beam splitter BS. In the first arm, the beam is projected by a spherical lens onto the plane of the CCD1 camera. The image of the beam cross-section at the beam waist

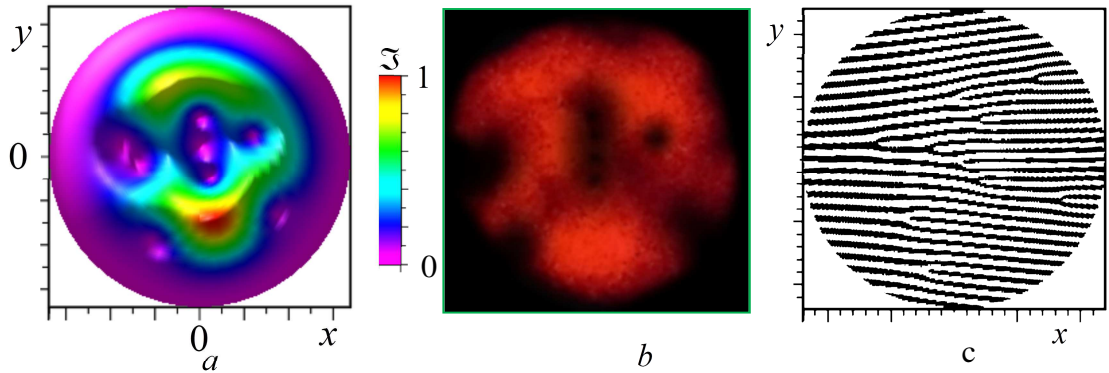

Fig. 2 A typical pattern of intensity distribution $\Im$ in a beam array with $N = 7$: (a) theory, (b) experiment and its holographic grating for its generation (c)

plane is processed by the computer that calculates the intensity moments $J_{p,q}$ in real time. In the process of computer processing, the intensity moments $J_{p,q}$ are substituted into the equations (4)-(8) and the squares of amplitudes $C_m^2$ and phases $\beta_m$ of the partial beams are monitored. For the calculation of the moments $J_{p,q}$, the photocurrent in each pixel of the image is multiplied by the cords $x^p \cdot y^q$, the results are summed over the entire image plane, normalized by the beam intensity $J_{00}$ and output to the computer monitor. In the second arm, the beam is focused by a cylindrical lens CL with a focal length $f_{cyl}$ at the plane of the CCD2 camera located at the focal plane of the lens CL. The image is processed by the computer in accordance with the method of

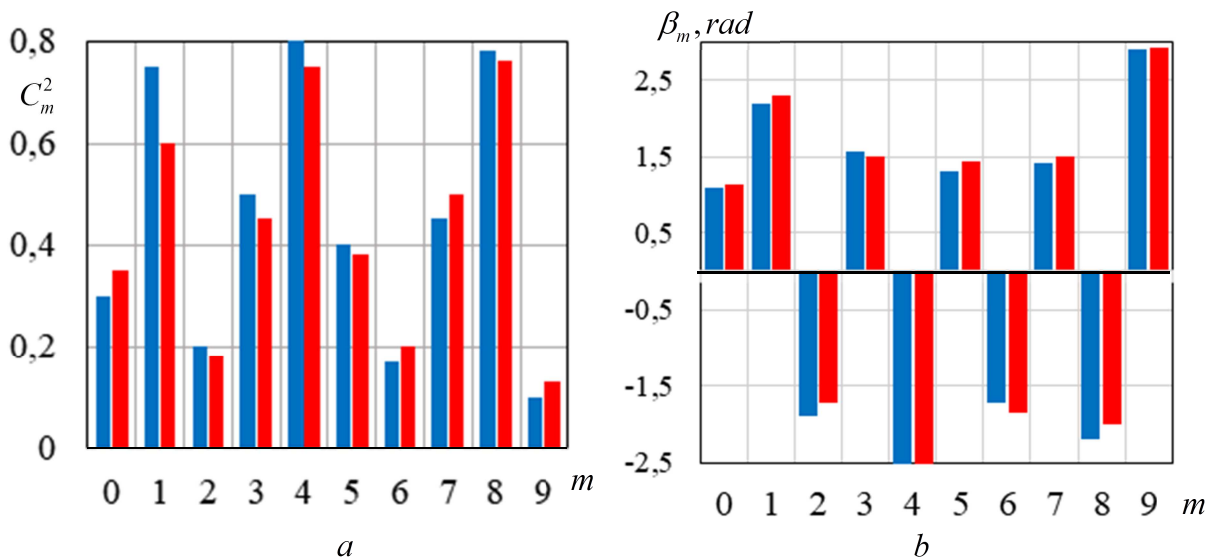

Fig.3 The vortex spectrum $C_m^2(m)$ (a) and (b) the phase histogram $\beta_m(m)$ of the array of $N = 10$ beams (blue – theory, red – experiment)

measuring the OAM per photon, described in detail in the paper [22]. The method was based on the measurement of the second-order intensity moments

To calibrate the optical systems in Fig,1 we first measured the squares of the amplitudes $C_m^2$ and phases $\beta_m$ of standard beam arrays with previously known parameters. First, we measured the waist radius of the initial beam $w_0$ at the observation plane via the second-order intensity moments [19] for normalizing the coordinates $x, y$. Then the computer program accidentally changed the magnitude of the amplitudes and phases. The obtained values were displayed on the monitor. One example of the intensity distribution $\Im(r, \varphi)$ of a beam array is shown in Fig. 2a,b. Fig.2c illustrates the diffraction grating for creating such beams. The vortex spectrum $C_m^2(m)$ and the phase distribution $\beta_m(m)$ are shown in Fig. 3. The red color in the figures corresponds to the obtained results, and the blue color shows the initial theoretical data. The measurement error for $N = 10$ beams did not exceed 3-4% for amplitudes and 5-6% for phases. We found out an increase in the measurement error for the phases with increasing a number of beams. So for N=20 beams, the phase measurement error increased to 10%, while the error of the square amplitude increase only to 5%. We associate such an increase of the measurement error with the low resolution of the liquid-crystal matrix in the light modulator (SLM) while in our case, the LQ-matrix resolution of the SLM is $900 \times 900$ *pixels*

**3.** Analyzing the dependence of the OAM per photon $\ell_z$ on the fractional topological charge $p$ of the singular beam (the OAM spectrum), Berry predicted [24] that the angular momentum $\ell_z(p)$ varies almost linearly (curve 1, Fig.4) with the growth of the $p$ value. The exact calculation of this case presented in Ref, [23] showed a nonmonotonic character of the OAM $\ell_z(p)$ (curce 2, Fig.4). As the topological charge $p$ rises, sharp bursts of the OAM occur at integer values $p = m$, while the width of the burst gets narrow.

In order to confirm or to refute these predictions, we have made measurements of the OAM $\ell_z(p)$ and the vortex spectrum $C_m^2(m)$. We measured the OAM for different fractional

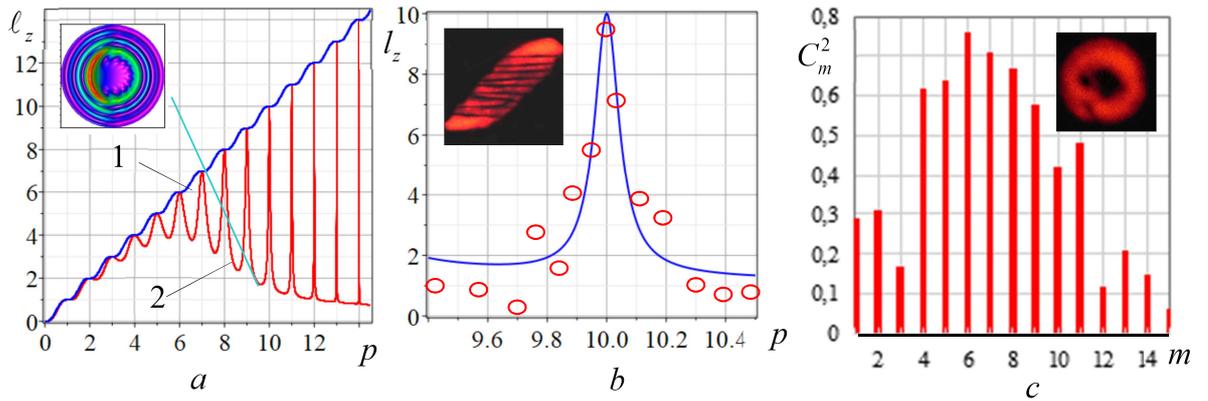

Fig.4 (a) Theoretical dependence of the OAM per photon $\ell_z$ on the fractional topological charge $p$: estimated [24] ( curve 1) and exact [23] (curve 2) calculations. (b) Experimental measurements of the OAM in the region $p = m = 10$ (circles) and theoretical solid curve. (c) Vortex spectrum $C_m^2(m)$ of the fractional topological charge $p = 9,5$ Callouts in figures: theoretical (a) and experimental (c) intensity distributions $\Im(r, \varphi)$ for $p = 9.5$ ; (b) experimental $\Im(r, \varphi)$ for $p = 9.7$ in the focal plane of a cylindrical lens with $f_{cyl} = 3.5\, sm$, the LQ-matrix resolution $900 \times 900$ *pixels* charges $p$ in region $p = m = 10$ using the second arm of the experimental setup in Fig. 1 (see Ref.[22]). The experimental circles in Fig. 4b are located near the theoretical curve indicating the sharp decrease of the OAM in the region of fractional charges $p$ in contrast to the predictions in Ref. [24]. But what physical processes are responsible for this effect? The question is answered by the experimental vortex spectrum $C_m^2(m)$, shown in Fig. 4c. We see that a small deviation from the integer-order charge $m$ leads to the birth of a wide range of vortices. Since the OAM per photon $\ell_z$ is calculated as the ratio of the beam OAM $L_z$ to its intensity $J_{00}$, for integer charges $m$ we deal only with one vortex, while a fractional charge $p$ corresponds to a large number of integer-order vortices with different weights. As a result, the OAM sharply decreases for fractional-order charges.

Thus, we have examined and tested the method of real-time measurement of the vortex spectrum (the nondegenerate case) in the composition of complex arrays of singular beams. The analysis of the degenerate case requires complementary measurements in the changed symmetry of the beam. The application of this approach e.g. for the terabit-information processing in optical fibers provides significant advantages in comparison with the traditional diffraction methods.

## References

1. G.J. Gbur. Singular optics. CRC Press, New York, 2017
2. Methods for Computer Design of Diffractive Optical Elements, Soifer, V.A., Ed., New York: John Wiley & Sons, Inc., 2002
3. A. Mair, A. Vaziri, G. Weihs, and A. Zeilinger, Entanglement of the orbital angular momentum states of photons, Nature 412, 313–316, (2001)
4. J. Leach, M. Padgett, S. Barnett, S. Franke-Arnold, and J. Courtial, "Measuring the orbital angular momentum of a single photon," Phys. Rev. Lett. 88(25), 257901 (2002)
5. Torres, J.P., Multiplexing twisted light, Nature Photonics, June 2012.
6. G. Gibson, J. Courtial, M. Padgett, M. Vasnetsov, V. Pas'ko, S. Barnett, and S. Franke-Arnold, Free-space information transfer using light beams carrying orbital angular momentum, Opt. Express 12, 5448{5456 (2004)
7. Khonina, S.N., Kazanskiy, N.L., and Soifer, V.A., Optical vortices in a fiber: mode division multiplexing and multimode selfimaging, Chapter in Recent Progress in Optical Fiber Research, Yasin, M., Harun, S.W., and Arof, H., Eds., INTECH publisher, Croatia, 2012
8. Khonina, S.N., Kotlyar, V.V., Soifer, V.A., Jefimovs, K., and Turunen, J., Generation and selection of laser beams represented by a superposition of two angular harmonics, J. Mod. Opt., 2004, vol. 51, pp. 761–773
9. V. A. Soifer, O. Korotkova, S. N. Khonina, E. A. Shchepakina. Vortex beams in turbulent media: review. Compuer Optics, 2016, 40, 605-624
10. M. S. Kirilenko, S. N. Khonina. Information transmission using optical vortices. Optical Memory and Neural Networks (Information Optics), 2013, 22, 81–89
11. J. Wang, J.-Y. Yang, I. M. Fazal, N. Ahmed, Y. Yan, H. Huang, Y.X. Ren, Y. Yue, S. Dolinar, M. Tur, and A.E. Willner, "Terabit free-space data transmission employing orbital angular momentum multiplexing," Nat. Photonics 6, 488–496 (2012)
12. J. Wang Advances in communications using optical vortices. Photon. Res. 4, B 14-28 2016
13. C. N. Alexeyev, Yu.A. Egorov, A.V. Volyar. Mutual transformations of fractional-order and integer-order optical vortices. Phys. Rev. A 96, 063807, 2017
14. C.N. Alexeyev, A.V. Volyar, M.A. Yavorsky. Energy transfer, orbital angular momentum, and discrete current in a double-ring fiber array, Phys. Rev. A 84, 063845, 2011

15. A. V. Volyar, Yu. A. Egorov. Super pulses of orbital angular momentum in fractional-order spiroid vortex beams. Optics Letters,. 43, 74-77 (2018)
16. M.P.J. Lavery, D. Robertson, G. C. G. Berkhout, G. Love and M. J. Padgett. Refractive elements for the measurement of the orbital angular momentum of a single photon" Opt. Express, 20, 3, (2012)
17. M. Sheikh, H. Ya. Rathore, S. Rehman. High-efficiency phase flattening based Laguerre–Gauss spectrometer using variable focus lenses. J. Opt, Soc. Am. B, 2017, 34, 76-81
18. M. K. Hu, ''Visual pattern recognition by moment invariants,'' IRE Transactions on 2 Information Theory, IT-8, 179-187, 1962
19. G.Nemes, A.Siegman, Measurement of all ten second-order moments of an astigmatic beam by the use of rotating simple astigmatic (anamorphic) optics, J. Opt. Soc. Am. A,11, 2257–2264, 1994
20. R. Procop, A. Reeves Survey of Moment-Based Techniques for Unoccluded Object Representation and Recognition. CVGIP: Graphical Models and Image Processing, 1992, 54(5): 438-460
21. A. Bekshaev,M. Soskin,M. Vasnetsov Optical vortex symmetry breakdown and decomposition ofthe orbital angular momentum of light beams. J. Opt. Soc. Am. A. – 2003. 20, 1635–1643
22. S. Alperin, R.. Niederriter, J. Gopinath, M. Siemens. Quantitative measurement of the orbital angular momentum of light with a single, stationary lens. Opt. Letters 41, 5019-5022 (2016)
23. T. Fadeyeva, A, Rubass, R. Aleksandrov, A, Volyar. Does the optical angular momentum change smoothly in fractional-charged vortex beams? J. Opt. Soc. Am. B 2014, 31 798-805
24. M.V. Berry Optical vortices evolving from helicoidal integer and fractional phase steps J. Opt. A: Pure Appl. Opt. 2004, 6, 259–68